\documentclass[12pt]{iopart}
\usepackage[utf8]{inputenc}
\usepackage[T1]{fontenc}

\expandafter\let\csname equation*\endcsname\relax
\expandafter\let\csname endequation*\endcsname\relax
\expandafter\let\csname thebibliography*\endcsname\relax
\usepackage{amsmath}

\usepackage{bm}
\usepackage{bbm} 
\usepackage{xcolor}
\usepackage{appendix}

\usepackage{graphicx}

\usepackage{geometry}
\newcommand\egaldef{\stackrel{\mbox{\upshape\tiny def}}{=}}
\newcommand{\hs}{\hat s}
\newcommand{\s}{\bm{s}}
\newcommand{\sdata}{{\underline s}}
\newcommand{\hsigma}{\hat{\sigma}}
\newcommand{\bsigma}{\bm{\sigma}}

\newcommand{\htheta}{\hat{\theta}}
\newcommand{\heta}{\hat{\eta}}

\newcommand{\DD}{\displaystyle}
\newcommand{\w}{W}
\newcommand{\uu}{U}
\newcommand{\uv}{V}
\newcommand{\fv}{{\bm \eta}}
\newcommand{\fh}{{\bm \theta}}
\newcommand{\var}{\bar{\sigma}}
\newcommand\1{\leavevmode\hbox{\rm \small1\kern-0.35em\normalsize1}}
\newcommand\ind[1]{\1_{\{#1\}}}

\usepackage{titlesec}

\begin{document}

\title{Gaussian-Spherical Restricted Boltzmann Machines}
\author{Aurélien Decelle, - Cyril Furtlehner}



\begin{abstract}
  We consider a special type of Restricted Boltzmann machine (RBM), namely a Gaussian-spherical RBM where the visible units have Gaussian priors while the vector of hidden variables
  is constrained to stay on an ${\mathbbm L}_2$ sphere. The spherical constraint having the advantage to admit exact asymptotic treatments, various scaling regimes are explicitly identified  
  based solely on the spectral properties of the coupling matrix (also called weight matrix of the RBM). Incidentally
  these happen to be formally related to similar scaling behaviours obtained in a different context dealing with spatial condensation of zero range processes. 
  More specifically, when the spectrum of the coupling matrix is doubly degenerated an exact treatment can be proposed to deal with finite size effects.
  Interestingly the known parallel between the ferromagnetic transition of the spherical model and the Bose-Einstein condensation can be made explicit in that case.
  More importantly this gives us the ability to extract all needed response functions with arbitrary precision for the training algorithm of the RBM. This 
  allows us then to numerically integrate the dynamics of the spectrum of the weight matrix during learning in a precise way. 
  This dynamics reveals in particular a sequential emergence of modes from the Marchenko-Pastur  bulk of singular vectors of the coupling matrix. 
\end{abstract}

\section{Introduction}

In the last decade, the field of machine learning became the center of attention of both the public domain and of the scientific research.
With the development of deep neural networks taking advantage of the GPU technology, the performance on classification tasks started to outperform human level at image recognition, and more recently,
generative model such as generative adversarial network~\cite{goodfellow2014generative} (GAN) have been able to generate images that cannot be distinguish from a true one~\cite{karras2019style}.
Despite recent significant advances~\cite{ShTi,JaGaHo} the theoretical understanding of deep learning lag behind these progresses, in various respects
like for instance on the interplay between adequate network architecture and complexity of the data. 

Statistical physics has been helpful in the past to clarify the learning process on idealized inference problems. In the 80',
before the A.I. winter, many works on neural networks were proposing some elements of understanding in terms of the theoretical phase diagram of some models.
For instance, a retrieval phase for the Hopfield model was determined along with the number of  patterns that can be retrieven in that case~\cite{Hopfield,AmGuSo1,AmGuSo2,AmGuSo3}.
Another example deals with the perceptron where again, the capacity for storing synthetic dataset can be computed \cite{gardner1988space,Derrida-Gardner,huang2013entropy}.
The storage of information in layered neural networks was also analyzed in~\cite{DoMe} with mean-field techniques. These approaches could then be adapted in many different contexts,
such as community detection~\cite{decelle2011inference}, compressed sensing~\cite{krzakala2012statistical} or traffic inference~\cite{FuLaAu} to mention only a few of them.
Typically in this kind of approach, the formalism of statistical physics relates the behaviour of the model to its position on a phase diagram in the large $N$ limit,
mean-field equations being used to characterize the free energy landscape and to sample efficiently the system.

In this work a somewhat similar path is followed to study generative models, by focusing on a ``tractable'' version of the restricted Boltzmann machine (RBM).
While RBM is considered to be a basic tool of machine learning, introduced more than 30 years ago~\cite{Smolensky}, it is still attracting a lot of interest,
both from the machine learning and the statistical physics communities. 
First, it is a model that can be handled without the need of GPU and can be run on a ordinary computer in reasonable time while solving non-trivial tasks.
Second, it has only one hidden layer in its classical formulation which allows the possibility to get some understanding of the learned hidden
features since they are directly linked to the visible variables.
Finally, it can be expressed as an Ising model and therefore, many standard tools developed by the statistical physics community can be used to determine its properties.

Originally, the RBM played an important role in deep learning as a way  to pre-train deep auto-encoders layerwise~\cite{HiSa}.
It is also in principle possible to stack many RBM to form a multi-layer generative model known as a Deep Boltzmann Machine (DBM)~\cite{salakhutdinov2009deep}.
Within the recent years, RBM has continuously attracted the interest of the research community, firstly because it can be easily used for both continuous and discrete
variables~\cite{krizhevsky2009learning,MuTa,cho2011improved,yamashita2014bernoulli} and the activation can be tuned to be either binary of relu~\cite{nair2010rectified};
secondly because for datasets of modest size it is able to deliver good results~\cite{hjelm2014restricted,hu2018latent}
comparable to the ones obtain from more elaborated network such as GAN (see for instance~\cite{yelmen2019creating}).
However, even for such a simple model, the learning procedure (what is learned, and how it is learned) is still very difficult to analyze with non-linear activation functions, 
in order to identify the key features and mechanisms
allowing it to work properly. Even for practical purpose, it is intrinsically difficult to efficiently estimate numerically the gradient w.r.t. the parameters of the model,
as soon as the network has learned non trivial modes. Empirical procedures have been proposed, first the contrastive divergence~\cite{Hinton_CD} (CD) and
the refined Persistence CD~\cite{Tieleman} (PCD) and later on a mean-field estimate~\cite{TAP_train},
none of these being fully satisfactory (see e.g.~\cite{tubiana2018restricted} for a more detailed discussion),
especially if one is willing to learn an empirical distribution with good accuracy.
For that purpose, recent works~\cite{barra2017phase,barra2018phase} using the analogy between the RBM and the Hopfield model
characterize the retrieval capacity of RBMs. RBM with sparse weight matrix have been considered in~\cite{TuMo} to analyze compositional mechanisms of features to create complex
patterns. Other works have focused on a mean-field theory for the RBMs, first to approximate the gradient and second to probe the mean-field landscape in
the general case~\cite{TAP_train,Mezard} or in the spherical case \cite{maillard2019high} or even to compute the entropy in a very simple case~\cite{huang2017statistical}.
Recently we also used a mean-field approach to understand the phase diagram of the model as a function of the spectrum of the weight matrix~\cite{DeFiFu,DeFiFu2}. We 
characterized in some way how the singular modes of the weight matrix evolve and  interact during learning, bringing forward a clustering model interpretation of the RBM
in terms of mean-field fixed points.

In this paper we study an  RBM with continuous symmetry, consisting of one layer of Gaussian variables (the visible one) and one layer of real variables with a spherical constraint. 
In the spirit of the original spherical Ising model introduced by Berlin and Kac~\cite{BerKac}, this offers the possibility to
say something relevant to the original model, by solving a simpler one.
It turns out that in a special setting the thermodynamical properties of the Gaussian-spherical RBM 
can be obtained exactly. This allows one to  devise an exact gradient ascent of the likelihood to learn the model, despite the fact that this model as we shall see
is able to encode only rather specific data.
The observation made previously on the spectral dynamics of the learning procedure~\cite{DeFiFu,DeFiFu2}, in particular that the modes of the weight matrix are learned by order of importance,
will be illustrated by an exact integration of the dynamical equations introduced in these works. In addition this solution could constitute a possibility to assess
approximate mean-field methods and empirical learning strategies.
\begin{figure}[ht]
\centerline{\resizebox*{0.7\textwidth}{!}{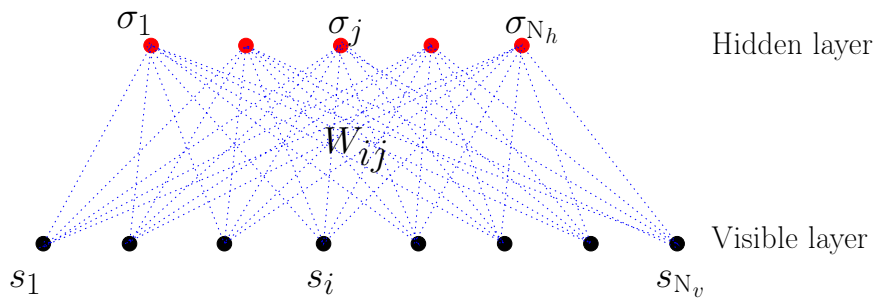}}
\caption{\label{fig:rbm} bipartite structure of the RBM.}
\end{figure}
The paper is organized as follows. In Section \ref{sec:def} we define the RBM with a spherical hidden layer and derive the likelihood and the response functions
exactly in the case of Gaussian visible units, together with the dominant behavior in the thermodynamic limit. In Section \ref{sec:asymp} we specify some properties
specific to the spherical constraint, as the way is occuring the onset of ferromagnetic order, the critical behavior of the magnetization associated to 
mode condensation, remarking and exploiting in passing some connection with spatial condensation in particle processes explored in~\cite{EvMaZi}.
Next,  the Section \ref{sec:exactA} focuses on a particular case where the spectrum of the weight matrix is doubly-degenerated and allows one to compute exactly
for finite size systems the partition function of the system.
Finally, in the section \ref{sec:dyn}  we exploit these results to  numerically integrate the spectral dynamics of the weight matrix during learning.

\section{Model definition} \label{sec:def}
\subsection{Boltzmann measure and associated likelihood}
The basic structure of the RBM is shown on Figure~\ref{fig:rbm}. It is a bipartite model connecting one layer of visible variables to one layer of hidden variables, these ones 
acting as a field to generate interactions among visible variables.
We define the visible variables $\{ s_i \}_{i=1, ..., N_v}$ and the hidden variables $\{ \sigma_i \}_{i=1, ..., N_h}$ both real valued, where $N_v$ 
and $N_h$ denotes the number of visible (resp. hidden) variables. $L = \sqrt{N_v N_h}$ will represent the size of the system and $\kappa = \sqrt{\frac{N_h}{N_v}}$
its shape. We define the energy function by

\begin{equation}
  E( \bm{s},\bm{\sigma} ) = - \sum_{i,j} w_{ij} s_i \sigma_j +\sum_i\frac{s_i^2}{2} - \sum_i \eta_i s_i - \sum_j \theta_j \sigma_j\label{eq:Erbm},
\end{equation}

\noindent $\w$ is the weight matrix between the visible and hidden variables, $\fv$ and $\fh$ are local fields exerted on variables. In this form  
the visible variables have a Gaussian prior ${\cal N}(0,1)$ in absence of hidden variables. 
The spherical constraint imposes an additional prior distribution on the hidden variables. Overall the distribution over $\bm{s}$ and $\bm{\sigma}$ is defined as

\begin{equation}\label{eq:prbm}
  p(\bm{s},\bm{\sigma}) = \frac{1}{Z}e^{-E( \bm{s},\bm{\sigma})}\ \delta\bigl( \sum_j \sigma_j^2 - \var^2 L \bigr)
\end{equation}

\noindent where $Z$ is the normalization factor and $\var$ a parameter of the model. In this setting, it is possible to 
diagonalize the distribution by using the singular value decomposition (SVD) of
the matrix $\w$:
\[
  w_{ij} = \sum_\alpha w_\alpha u_i^\alpha v_j^\alpha
\]
where $u^\alpha$ are left singular vectors, attached to the visible space, while $v^\alpha$ are right singular vectors attached to the hidden space and $w_\alpha$
are the singular values. Depending on whether $N_v>N_h$ or $N_v<N_h$ the set $u^\alpha$ or the set $v^\alpha$ is not a complete orthonormal set of respectively
the visible or the hidden space. If we assume for instance that $N_h < N_v$ the matrix corresponding to the left singular vectors has to be complemented by 
$N_v-N_h$ arbitrary orthonormal vectors to form a complete basis of the visible space. For the moment we don't need to specify whether $N_v$ is larger than $N_h$, 
denote $N = \min\{N_v,N_h\}$ and assume that $\uu$ and $\uv$ represent complete basis respectively of the visible and hidden space.

The joint distribution~(\ref{eq:prbm}) is conveniently expressed  by means of the components of the visible and hidden vectors in these bases:
\[
  \hs_\alpha = \frac{1}{\sqrt{L} } \sum_{i=1}^{N_v} U_{\alpha i} s_i\qquad
  \hat{\eta}_\alpha =  \frac{1}{\sqrt{L}} \sum_{i=1}^{N_v}  U_{\alpha i}\eta_i \qquad
\]
for $\alpha\in\{1,\ldots N_v\}$ and
\[
  \hsigma_\alpha = \frac{1}{\sqrt{L}} \sum_{j=1}^{N_h} V_{\alpha j} \sigma_j\qquad
  \hat{\theta}_\alpha =  \frac{1}{\sqrt{L}} \sum_{j=1}^{N_h}  V_{\alpha j}\theta_j 
\]
for $\alpha\in\{1,\ldots N_h\}$. These obey the following normalization rules:
\[
\sum_{i=1}^{N_v} s_i^2 = L \sum_{\alpha=1}^{N_v} \hs_\alpha^2 \qquad
\sum_{j=1}^{N_h} \sigma_j^2 = L \sum_{\alpha=1}^{N_h} \hsigma_\alpha^2 \qquad
\sum_{i=1}^{N_v} \eta_i s_i = L \sum_{\alpha=1}^{N_v}  \hat{\eta}_\alpha\hs_\alpha \qquad
\sum_{j=1}^{N_h} \theta_i \sigma_i = L \sum_{\alpha=1}^{N_h}  \hat{\theta}_\alpha\hsigma_\alpha 
\]
we obtain
\begin{align*}
  p({\bf \hs},{\bm \hsigma}) = \frac{1}{Z} \exp\left( L \sum_{\alpha=1}^N \left[ w_\alpha \hs_\alpha \hsigma_\alpha + \heta_\alpha \hs_\alpha 
+ \htheta_\alpha \hsigma_\alpha \right] - L\sum_{\alpha=1}^{N_v} \frac{\hs_\alpha^2}{2} \right) 
\delta\Big(L\sum_{\alpha=1}^{N_h} \hsigma_\alpha^2 - \var^2L\Big).
\end{align*}
From these transformations, we expect  $\hs_\alpha, \hsigma_\alpha$ and also $\htheta_\alpha,\heta_\alpha$ to scale like $\sim L^{-0.5}$.
In this representation the SVD modes are coupled by the spherical constraint. 
To get the distribution of the visible variables alone, we have to integrate over the hidden variables which can be done first 
by using the Fourier representation of the $\delta$ function
\begin{equation}\label{eq:prbm2}
p(\bm{\hs},\bm{\hsigma}) = \frac{1}{2i\pi Z}\int_{a-i\infty}^{a+i\infty}dz \exp\left(L 
\Bigl( \sum_\alpha\left[ w_\alpha \hs_\alpha \hsigma_\alpha + \heta_\alpha \hs_\alpha 
+ \htheta_\alpha \hsigma_\alpha \right]
- \sum_\alpha\frac{\hs^2_\alpha}{2} - z \bigl(\sum_\alpha \hsigma_\alpha^2 - \var^2\bigr) \Bigr)\right),
\end{equation}
with $a>0$.
With the change of variable $z' = 2\var z/\Sigma(\bm{\hs})$ we get
\[
p(\bm{\hs}) = \frac{1}{2iLZ} \Bigl(\frac{2\pi}{L\var\Sigma(\bm{\hs})}\Bigr)^{N_h/2-1} \exp\left(L\sum_{\alpha=1}^{N_v}\bigl(\heta_\alpha \hs_\alpha-\frac{\hs_\alpha^2}{2}\bigr)\right)
\int_{a-i\infty}^{a+i\infty}\frac{dz}{z^{N_h/2}}\exp\left(\frac{L\var\Sigma(\bm{\hs})}{2}\bigl(z+\frac{1}{z}\bigr)\right)
\]
with
\[
\Sigma^2(\bm{\hs}) \egaldef \sum_{\alpha=1}^N(w_\alpha\hs_\alpha+\htheta_\alpha)^2.
\]
The integration over $z$ can actually be rewritten as (for $N_h\ge 2$)
\begin{equation}\label{eq:ps}
p(\bm{\hs}) = \frac{(2\pi)^{N_h/2}}{2LZ}\tilde I_{N_h/2-1}\bigl(L\var\Sigma(\bm{\hs})\bigr)
\exp\left(L\sum_{\alpha=1}^{N_v}\bigl(\heta_\alpha \hs_\alpha-\frac{\hs_\alpha^2}{2}\bigr)\right),
\end{equation}
where  
\[
\tilde I_\nu(x) = x^{-\nu}I_\nu(x),
\]
with  the modified Bessel function
\[
I_\nu(x) = \frac{x^\nu}{2}\sum_{k=0}^{+\infty}\frac{\bigl(\frac{x}{2}\bigr)^{2k}}{k!\Gamma(\nu+k+1)}.
\]
The partition function is also given by means of a single integral after integrating over visible and hidden variables the form~(\ref{eq:prbm2}) after the change $z'=2z$:
\begin{equation}\label{eq:Zcontour}
Z = \frac{1}{2i\pi}\int_{a-i\infty}^{a+i\infty}dz e^{L\phi(z)}, 
\end{equation}
where
\begin{equation}\label{eq:phi}
\phi(z) = \frac{\var^2 z}{2}-\frac{\delta}{2}\log(z)+\frac{h_0^2}{2z}+\frac{1}{2}\sum_{\alpha=1}^{N_v}\eta_\alpha^2
+\frac{1}{2}\sum_{\alpha=1}^{N}\Bigl[\frac{h_\alpha^2}{z-w_\alpha^2}-\frac{1}{L}\log(z-w_\alpha^2)\Bigr],
\end{equation}
up to a constant and
\begin{align*}
\delta &\egaldef (\kappa-\kappa^{-1})\ind{N_h>N_v}, \\  
h_0^2 &\egaldef \ind{N_h>N_v}\sum_{\alpha=N_v+1}^{N_h} \htheta_\alpha^2,\\
h_\alpha &\egaldef \heta_\alpha w_\alpha + \htheta_\alpha.
\end{align*}

\subsection{Learning algorithm}\label{sec:LA}
The objective of the standard learning procedure of the RBM is to find the set of parameters $\{\w,\fv,\fh\}$ such that the likelihood of a 
given dataset $\sdata$ be maximal. This is done by conventional gradient ascent of the log likelihood (LL).
The conventional gradient of the LL w.r.t. the parameters is given by
\begin{align*}
\frac{\partial {\cal L}}{\partial w_{ij}} &= \langle s_i\sigma_j p(\bsigma\vert \s)\rangle_{\rm Data}-\langle s_i\sigma_j\rangle_{\rm RBM}\\[0.2cm]
\frac{\partial {\cal L}}{\partial \eta_i} &= \langle s_i\rangle_{\rm Data} -  \langle s_i\rangle_{\rm RBM} \\[0.2cm]
\frac{\partial {\cal L}}{\partial \theta_j} &=\langle \sigma_j p(\bsigma\vert \s)\rangle_{\rm Data} - \langle \sigma_j\rangle_{\rm RBM}
\end{align*}
This requires to compute various response functions of the RBM and the conditional probability $p(\bsigma\vert \s)$. 
As shown in~\cite{DeFiFu,DeFiFu2} it is convenient to rewrite the gradient in the frame defined by the SVD modes of the weight matrix.
As already seen it is here especially adapted since the RBM measure is naturally expressed in this frame.
In addition, the specificity of the Gaussian-spherical model is that the joint distribution of the visible variables~(\ref{eq:ps}) is invariant w.r.t. a rotation
of the singular vector $\uv$. This means that we can use a more economical gradient. In addition
to modifications of $\{w_\alpha,\heta_\alpha,\htheta_\alpha\}$ we are led to consider infinitesimal rotation $\Omega_{\alpha\beta}$ between modes $\alpha$ and $\beta$
of the visible SVD basis only. Here $\Omega_{\alpha\beta}$ is a skew-symmetric operator corresponding to the change 
\begin{align*}
du_\alpha &= \Omega_{\alpha\beta} u_\beta,\\[0.2cm]
du_\beta  &= -\Omega_{\alpha\beta} u_\alpha.
\end{align*}
Our simplified LL gradient now reads:
\begin{align*}
\frac{1}{L}\frac{\partial {\cal L}}{\partial w_\alpha} &= \langle\hs_\alpha\hsigma_\alpha p(\bm{\hsigma}\vert \bm{\hs})\rangle_{\rm Data}-\langle\hs_\alpha\hsigma_\alpha\rangle_{\rm RBM}\\[0.2cm]
\frac{1}{L}\frac{\partial {\cal L}}{\partial \heta_\alpha} &= \langle\hs_\alpha\rangle_{\rm Data}-\langle\hs_\alpha\rangle_{\rm RBM}\\[0.2cm]
\frac{1}{L}\frac{\partial {\cal L}}{\partial \htheta_\alpha} &= \langle \hsigma_\alpha p(\bm{\hsigma}\vert \bm{\hs})\rangle_{\rm Data}-\langle \hsigma_\alpha\rangle_{\rm RBM}\\[0.2cm]
\frac{1}{L}\frac{\partial {\cal L}}{\partial \Omega_{\alpha\beta}} &= \langle (w_\alpha\hs_\alpha\hsigma_\beta-w_\beta \hs_\beta\hsigma_\alpha) p(\bm{\hsigma}\vert \bm{\hs})\rangle_{\rm Data}
-\langle (w_\alpha\hs_\alpha\hsigma_\beta-w_\beta \hs_\beta\hsigma_\alpha)\rangle_{\rm RBM}
\end{align*}
with 
\[
p(\bm{\hsigma}\vert \bm{\hs}) = \frac{\var^2L}{(2\pi)^{N_h/2}}
\frac{\exp\Bigl(L \sum_\alpha w_\alpha \hs_\alpha \hsigma_\alpha + \htheta_\alpha \hsigma_\alpha\Bigr)}{ \tilde I_{N_h/2-1}\bigl(L\var\Sigma(\bm{\hs})\bigr)}
\delta\Bigl(L\sum_\alpha \hsigma_\alpha^2 -\var^2 L\Bigr).
\]
This results in the following (continuous) update equations for the parameters and the dataset ${\cal S}$:
\begin{align}
\frac{dw_\alpha}{dt} &= \langle\hs_\alpha\hsigma_\alpha p(\bm{\hsigma}\vert \bm{\hs})\rangle_{\rm Data}-\langle\hs_\alpha\hsigma_\alpha\rangle_{\rm RBM}, \label{eq:wa}\\[0.2cm]
\frac{d\heta_\alpha}{dt} &= \langle\hs_\alpha\rangle_{\rm Data}-\langle\hs_\alpha\rangle_{\rm RBM}-\sum_\beta \Omega_{\alpha\beta}\heta_\beta, \label{eq:eta}\\[0.2cm]
\frac{d\htheta_\alpha}{dt} &= \langle \hsigma_\alpha p(\bm{\hsigma}\vert \bm{\hs})\rangle_{\rm Data}-\langle \hsigma_\alpha\rangle_{\rm RBM}, \label{eq:htheta}\\[0.2cm]
\frac{d\hs_\alpha^k}{dt} &= -\sum_\beta \Omega_{\alpha\beta}\hs_\beta^k,\qquad\forall k\in{\cal S},\label{eq:hsa}
\end{align}
with 
\begin{equation}\label{eq:omega}
\Omega_{\alpha\beta} = \langle (w_\alpha\hs_\alpha\hsigma_\beta-w_\beta \hs_\beta\hsigma_\alpha) p(\bm{\hsigma}\vert \bm{\hs})\rangle_{\rm Data}
-\langle (w_\alpha\hs_\alpha\hsigma_\beta-w_\beta \hs_\beta\hsigma_\alpha)\rangle_{\rm RBM}
\end{equation}
and 
where $dt/L$ represents the learning rate. Here the last update equation corresponds to simply adapting the data projection on the rotated basis.
Note that the same is done also for the second update equation concerning the field projection $\heta$, which is optional here 
but coherent with the conventional update rules and useful in practice.
Note also that the singularity of the conventional gradient observed in~\cite{DeFiFu} for pairs of modes with identical singular values has disappeared. 
Hence, computing the gradient requires to evaluate one and two-points correlation functions of SVD variables. 

As seen in the previous section, the LL takes the form
\[
  {\cal L} = \Big\langle\log\Bigl(\tilde I_{N_h/2-1}\Bigl[L\var\Sigma(\bm{\hs})\Bigr]\Bigr) +
L\sum_\alpha\bigl(\heta_\alpha \hs_\alpha-\frac{\hs_\alpha^2}{2}\bigr)\Big\rangle_\text{\rm Data}
-\log\bigl(Z\bigr).
\]
Compared to the simple Gaussian RBM likelihood, we see one important difference: eigenvalues of $\w$ do interact, in particular  
via the empirical term which is now 
a nonlinear, monotonically increasing function of $\Sigma(\bm{\hs})$.
With help of the identity
\[
\frac{d\tilde I_\nu(x)}{dx} = x\tilde I_{\nu+1}(x),
\]
we get from this the gradient of the log likelihood in the form: 
\[
\frac{\partial {\cal L}}{\partial w_a} = \var\Big\langle \hs_\alpha(w_\alpha\hs_\alpha+\htheta_\alpha)
\frac{I_{N_h/2}\bigl(L\var\Sigma(\bm{\hs})\bigr)}{\Sigma(\bm{\hs}) I_{N_h/2-1}\bigl(L\var\Sigma(\bm{\hs})\bigr)}\Big\rangle_\text{Data}-\frac{\partial\log(Z)}{\partial w_\alpha}
\]
Using the following asymptotic expression for large $\nu$ (see e.g.~\cite{AbSt})
\[
I_\nu(\nu z) \sim \frac{1}{\sqrt{2\pi\nu}}\frac{e^{\nu\eta}}{(1+z^2)^{1/4}}\qquad\text{with}\qquad \eta = \sqrt{1+z^2}+\log\frac{z}{1+\sqrt{1+z^2}},
\]
resulting from a saddle point approximation of the modified Bessel function, 
we obtain the asymptotic expression
\[
\langle\hs_\alpha\hsigma_\beta p(\bm{\hsigma}\vert \bm{\hs})\rangle_{\rm Data}  = \var
\left\langle \hs_\alpha\frac{w_\beta  \hs_\beta +\htheta_\beta}{1+\sqrt{1+\var^2\Sigma(\bm{\hs})^2}}\right\rangle_{\rm Data},
\]
valid for large $L$.

The remaining point to address now in order to be able to train such a machine is the estimation of the partition function and its derivatives. 
For the rest of the paper, the local fields $\heta_\alpha$ on the visible variables will be set to zero to lighten the presentation.

\section{Thermodynamical properties}\label{sec:asymp}
The expression of the partition function given by eq.~(\ref{eq:Zcontour}-\ref{eq:phi}) indicates that the physical properties of the Gaussian-spherical RBM depend only on the spectrum
of its weight matrix in absence of the fields as for the ordinary spherical model~\cite{Pastur}. 
Standard treatments of the spherical model (see \cite{BerKac,KoThJo,Baxter}) rely on a saddle point approximation of the
contour integral representation of $Z$ given by eq.~(\ref{eq:Zcontour}). Here we recall and straightforwardly adapt these arguments to our needs by making simple assumptions on
the limit spectrum of $W$ when $L\to\infty$. 
This leads us in a second step to establish incidentally a connections with condensation phase transition analyzed in the context of factorized steady states~\cite{EvMaZi}.
\subsection{Ferromagnetic transition}
\begin{figure}[ht]
\centerline{\resizebox*{0.9\textwidth}{!}{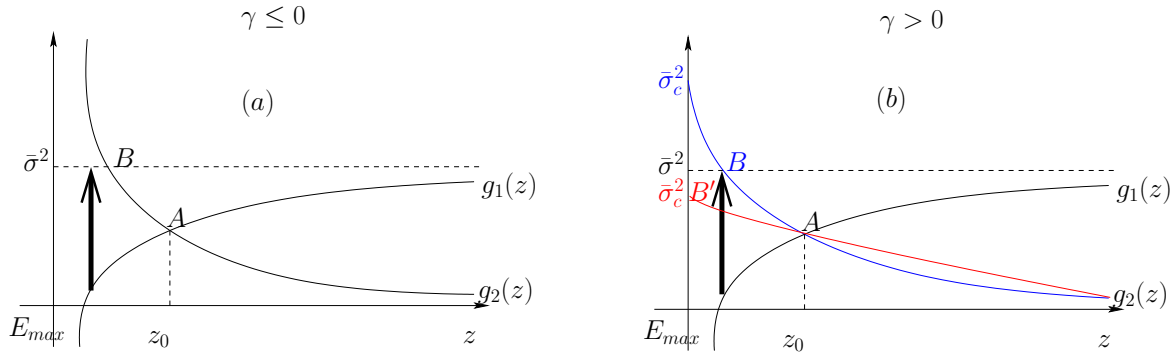}}
\caption{\label{fig:saddle_point} Various scenarios for the saddle points solution $z_0$. Point $A$ corresponds to the intersection between $g_1$ and $g_2$ 
at finite field $h(E)$, the black arrow the movement of $g_1$ when the fields tend to zero while 
points $B$ and $B'$ corresponds to the limit intersection at vanishing fields when respectively $\var<\var_c$ (in blue) and $\var>\var_c$ (in red).}
\end{figure}

\noindent First notice that $\phi$ given in (\ref{eq:phi}) is convex on the domain of interest:
\[
\phi''(z) = \frac{\delta}{2z^3}+\frac{h_0^2}{z^3}+\sum_\alpha\Bigl[\frac{h_\alpha^2}{(z-w_\alpha^2)^3}+\frac{1}{2L}\frac{1}{(z-w_\alpha^2)^2}\Bigr]
  > 0,\qquad \text{for}\ z>w_{\rm max}^2,
\]
with $w_{\rm max}$ the highest singular value, so there is only one solution $z_0$ to the saddle point equation
allowing for the following approximation:
\[
Z \sim_{L\to\infty} \frac{\exp\bigl(L\phi(z_0)\bigr)}{\sqrt{2\pi L\vert \phi''(z_0)\vert}}.
\]
At the saddle point the free energy per degree of freedom is given by 
\[
f = -\phi(z_0,\eta,\theta).
\]
From these quantities we can in principle get all the needed response functions (see Appendix~\ref{app:resp}). We will focus here on the computation of the spontaneous magnetizations as a function of the spectrum of $\w$. Their expressions can be obtain using
\begin{align}
\langle\hs_\alpha\rangle &= -\frac{\partial f}{\partial \heta_\alpha} = w_\alpha\frac{h_\alpha}{z_0-w_\alpha^2},\label{eq:ms} \\[0.2cm]
\langle \hsigma_\alpha\rangle &= -\frac{\partial f}{\partial \htheta_\alpha} = \frac{h_\alpha}{z_0-w_\alpha^2}.\label{eq:msigma}
\end{align}
This gives us relations between magnetizations and $z_0$.
\noindent In order to analyze further the thermodynamic properties of the system some assumptions have to be made on the spectral properties of $\w$.
Let us define the spectral density (SD) associated to $WW^T$:
\[
\rho_L(E) \egaldef \frac{1}{L}\sum_{\alpha=1}^{N_h}\delta\bigl(E-w_\alpha^2\bigr),
\]
(which includes zero modes $w_\alpha=0$ for $\alpha>N_v$ whenever $N_h>N_v$).
In the thermodynamic limit it is assumed that the SD tends to a well defined limit distribution
\[
\lim_{L\to\infty} \rho_L(E) = \rho(E).
\]
This leads to
\[
\phi(z) = \frac{1}{2}\var^2z+\frac{1}{2}\int_0^{E_{\rm max}}dE \rho(E)\Bigl(\frac{h(E)^2}{z-E}-\log(z-E)\Bigr),
\]
with some upper bound $E_{\rm max}$ of the SD, and where $h(E)$ is any smooth function taking the value $\sqrt{L}h_\alpha$ for $E=w_\alpha^2$ at finite $L$, $h_\alpha$ 
being expected to be $O(1/\sqrt{L})$ in general. $\rho(E)h(E)^2$ represents the SD of the external field.
As in~\cite{KoThJo} for instance, we have to distinguish between a situation where the SD has isolated dominant modes and the situation with just a continuous bulk of modes
bounded by $E_{\rm max}$. 
When the SD has no isolated dominant mode we look for a solution $z_0$ to the saddle point equation
\begin{align*}
\phi'(z) &= \frac{1}{2}\bigl(g_1(z)-g_2(z)\bigr) \\[0.2cm]  
&=0
\end{align*} 
where 
\begin{align*}
g_1(z) &\egaldef \var^2-\int_0^{E_{\rm max}} dE \frac{\rho(E)}{(z-E)^2} h(E)^2,\\[0.2cm]
g_2(z) &\egaldef \int_0^{E_{\rm max}} dE \frac{\rho(E)}{z-E},
\end{align*}
Let us call
\begin{equation}
\var^2_c \egaldef  g_2(E_{\rm max})  \label{eq:varc}
\end{equation}
The properties of the system depends on the behavior of $\rho(E)$ near $E_{\rm max}$.  
A thorough discussion of its influence on the physical properties of the spherical model can be found in~\cite{Pastur}. 
Here we restrict the discussion to behavior of the type $\rho(E)\sim (E_{\rm max}-E)^\gamma$ with the exponent $\gamma>-1$. This cover various study cases
like for instance $d$-dimensional regular lattices $\gamma = d/4-1$ or $\gamma=1/2$ for  i.i.d. random matrices. 
In order to get closed form expressions we shall consider 
the following beta distributions for the SD:  
\begin{align}
\rho(E) &=\frac{\kappa}{B(1-\gamma,\gamma+1)}\frac{E^{-\gamma}(E_\text{\rm max}-E)^\gamma}{E_{\rm max}}  \qquad\text{with}\qquad \gamma\in]-1,1[,\label{eq:sd}\\[0.2cm]
\rho(E)h(E)^2 &=  \frac{h^2}{B(\beta+1,1-\beta)}\frac{E^\beta(E_\text{\rm max}-E)^{-\beta}}{E_{\rm max}}  \qquad\text{with}\qquad \beta\in]-1,1[,\label{eq:h}
\end{align}
where the beta function takes here the special form
\[
B(1-\gamma,1+\gamma) = \frac{\gamma\pi}{\sin(\gamma\pi)},
\]
and with $h^2$ the squared norm of the external field. $\rho(E)h(E)^2$ represents the SD of the external fields. 
This setting can be useful to study the response function at the top of the spectrum, by simply letting $\beta\to 1$.
$\var_c$ is infinite for $\gamma\le 0$ and finite otherwise with 
\[
\var^2_c = \frac{1}{E_{\rm	 max}}\Bigl(\delta+\frac{\kappa}{\gamma}\Bigr),\qquad\gamma>0
\]
in the latter case.
The different scenarios for obtaining $z_0$ are sketched on Figure~\ref{fig:saddle_point}.
\begin{itemize}
\item When $\gamma \leq 0$, $g_2$ diverges close to $E_{\rm max}$ and therefore the intersection $A$ with $g_1$ always converges to the point $B=(z_0 > E_{\max},\var^2)$. In that case, 
there is no way that  $(z-E)$  goes to zero for $E\le E_{\rm max}$ when $h \rightarrow 0$ and therefore all the magnetizations~(\ref{eq:ms},\ref{eq:msigma}) vanish.
\item When $\gamma >0$ 
we have to distinguish between two cases. First, we consider condensation on modes that have $E< E_{\rm max}$ by applying small vanishing fields on these modes. 
In that case, since when $h \rightarrow 0$ we have  $z_0 \geq E_{\rm max}$, again the magnetization will simply vanish since the denominator in~(\ref{eq:ms},\ref{eq:msigma}) is finite and non-zero. 
For modes at $E_{\rm max}$ if $\var < \var_c$, then $z_0 \rightarrow E>E_{\rm max}$ when we put the field to zero 
therefore giving the same results as in the first scenario. Now, if instead $\var\ge \var_c$, $z_0 \rightarrow E_{\rm max}$ 
$h(E_{\rm max})\rho(E_{\rm max})/(z_0-E_{\rm max})$ has a finite limits given below. 
We obtain a spontaneous magnetization in that case. 
\end{itemize}

The spontaneous magnetization is obtained as follows for the cases where a saddle point solution exists.
If a field $h_\alpha=h_{\rm max}$ is concentrated on the largest mode, this is equivalent to choose $h(E)$ such that  
$\rho(E)h(E)^2 = h_{\rm max}^2\delta(E-E_{\rm max})$.
We get from the saddle point equation:
\[
\var^2 - \frac{h_{\rm max}^2}{(z_0(h_{\rm max})-E_{\rm max})^2} = \var^2_c.
\]
Now, when $z_0(h_{\max})\to E_{\rm max}$ and we let $h_{\rm max} \to 0$ we obtain
\[
\lim_{h_{\rm max}\to 0} \frac{h_{\rm max}}{z_0(h_{\max})-E_{\rm max}} = \sqrt{\var^2-\var^2_c}. 
\]
Eliminating $z_0-E$ in~(\ref{eq:ms},\ref{eq:msigma}) yields the spontaneous magnetization 
\begin{align*}
\langle\hs_\alpha \rangle &= w_\text{\rm max}\var\sqrt{\var^2-\var_c^2},\\[0.2cm]
\langle \hsigma_\alpha \rangle &= \sqrt{\var^2-\var_c^2},
\end{align*}
and
\[
\langle\hs_\alpha\hsigma_\beta \rangle = \delta_{\alpha\beta}w_\alpha(\var^2-\var_c^2).
\]
When the highest mode acquires a macroscopic magnetization in the ferromagnetic phase, the resulting distribution $p({\bm s})$ becomes bimodal along this direction. 
It is also worth noticing that when the highest mode is degenerated $n$ times, in absence of any external fields the system has an $O(n)$ symmetry
corresponding to rotations in the subspace defined by these vectors. This results in that case into a distribution concentrated on a $n$-dimensional sphere
in the ferromagnetic phase. The specific shape of the condensed distribution will be studied in the next subsection.

\subsection{Condensation mechanism in thermodynamic limits}\label{sec:condensate}
The scaling form~(\ref{eq:sd},\ref{eq:h}) of the SD of singular values and field densities allows us to make an explicit connection
with scaling function derived in a different context, namely condensation of factorized steady states~\cite{EvMaZi}.
After dropping irrelevant terms and making the change $z'=(z-E_{\rm max})/E_{\rm max}$ while absorbing $E_{\rm max}$ in the definition of $\var$ we get the following form of the partition
function:
\[
Z_{L,N}[\var,h] = \frac{1}{2i\pi}\int_{-i\infty}^{i\infty}dz e^{\frac{L}{2}\phi(z,\var,h)},
\]
with (see Appendix~\ref{app:scaling})
\begin{equation}\label{eq:phizvarh}
\phi(z,\var,h) = \var^2(z+1)-\frac{\kappa}{\gamma}\int_0^zdu\Bigl[1-\Bigl(\frac{u}{1+u}\Bigr)^\gamma\Bigl]\ +\ \frac{h^2}{\beta}\Bigl[\Bigl(\frac{1+z}{z}\Bigr)^\beta-1\Bigr].
\end{equation}
For large $L$ the rescaling $L^{1/(1+\gamma)}z \to z$  leads to the scaling behavior
\[
Z_{L,N}(\var,h) = e^{\frac{L}{2}\bigl(\var^2-\frac{h^2}{\beta}\bigr)}\left(L^{-\frac{1}{\gamma+1}}V_{\gamma,\beta}\Bigl(L^{\frac{\gamma}{1+\gamma}}(\var^2-\var_c^2),L^{\frac{2+\beta+\gamma}{1+\gamma}}h^2\Bigr) +{\cal O}\Bigl(\frac{1}{L^{\frac{2}{1+\gamma}}}\Bigr)\right),
\]
valid for $(\gamma,\beta)\in(]-1,0[\cup]0,1[)^2$,
where the following scaling function has been introduced
\begin{align*}
V_{\gamma,\beta} (x,y) &= \int_{-i\infty}^{i\infty} dz \exp\Bigl(\frac{1}{2}xz+bz^{\gamma+1}+c \frac{y}{z^{1+\beta}}\Bigr),\\[0.2cm]
&= \frac{1}{\pi} \int_0^\infty du\ e^{-b_2u^{\gamma+1}-c_2\frac{y}{u^{\beta+1}}}
\cos\Bigl(\frac{xu}{2}-b_1u^{\gamma+1}+c_1\frac{y}{u^{\beta+1}}\Bigr),
\end{align*}
with  
\begin{align*}
b=\frac{\kappa}{2\gamma(\gamma+1)}\qquad b_1 &= b\cos\bigl(\frac{\gamma\pi}{2}\bigr)\qquad b_2 = b\sin\bigl(\frac{\gamma\pi}{2}\bigr)\\[0.2cm] 
c=\frac{1}{2\beta}\ \ \qquad c_1 &= c\cos\bigl(\frac{\beta\pi}{2}\bigr)\qquad c_2 = c\sin\bigl(\frac{\beta\pi}{2}\bigr)
\end{align*}
and where the change of variable $z=\pm iu$ is used  for  $Im(z)\pm 0$. 
In~\cite{EvMaZi} the same scaling function (at $y=0$) is
encountered albeit in a different context. We can therefore closely follow their analysis to describe the transition to ferromagnetic order of the present spherical model. 
\begin{figure}[ht]
\centering
\includegraphics[scale=0.6]{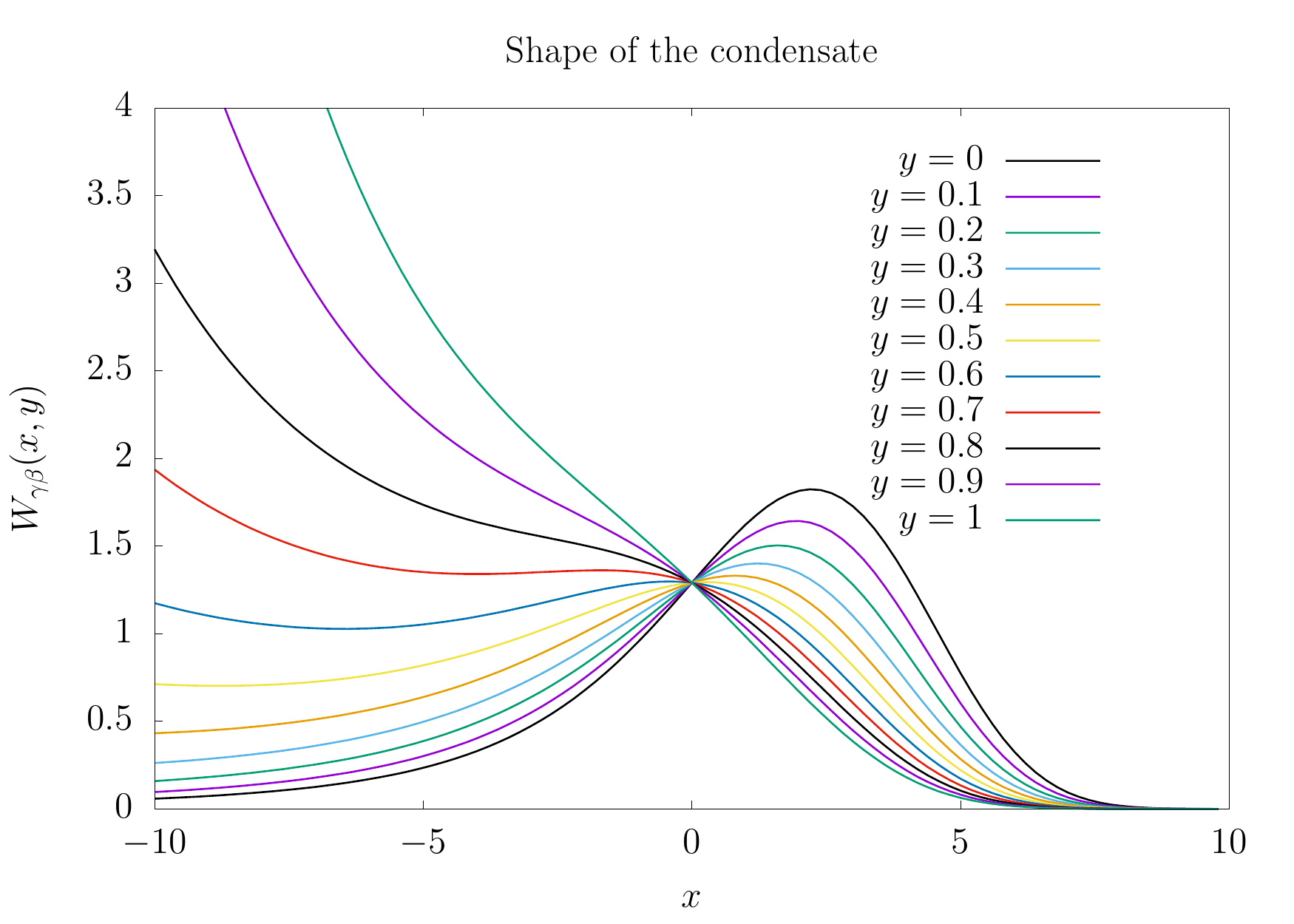}
\caption{Shape of the condensate distribution ($\gamma=0.5$) along a given mode $\alpha$ as the scaled  distance $y=L^{\frac{1}{1+\gamma}}(1-\epsilon_\alpha)$ from the upper boundary is increased,
  $x$ being the scaled fraction of variance $L^{\frac{\gamma}{1+\gamma}}(\sigma_\alpha^2-V_{\rm ex})$ along this mode. The bump in the distribution disappear for $y\ge y_{0.5}\simeq 0.72$, while the second
  mode correspond to the range $y\in[1.77,2.81[$ for $\gamma=0.5$.
  \label{fig:condensate}}
\end{figure}
As seen previously, when $\gamma>0$ there is a possibility for dominant modes to generate ferromagnetic order when $\var_c <\var$. 
In presence of an external field ($h^2>0$) there is always a solution to the saddle point equation and no transition occurs at $\var_c$.  Instead,
in absence of external fields and $\gamma>0$, there is no solution to the saddle point equation when $\var_c <\var$ while there is always one
in the opposite case, and the transition corresponds to the onset of ferromagnetic order materialized by condensation along the dominant modes. In that case a finite fraction 
of the overall variance of the distribution is captured by one or possibly a small number of modes. 

In order to study the condensate we need to express the marginal probabilities $p_{\hs_\alpha}(x)\egaldef P(\hs_\alpha = x)$ and $p_{\hsigma_\alpha}(x) \egaldef P(\hsigma_\alpha=x)$.
For any given mode $\alpha$ we have (in absence of external fields)
\begin{align*}
p_{\hs_\alpha}(x) &= \int dy p_{\hsigma_\alpha} (y) e^{L \bigl(\sqrt{\epsilon}_\alpha xy -\frac{x^2}{2}\bigr)} \\[0.2cm]
p_{\hsigma_\alpha} (x) &= \frac{Z_{L,N-1}\bigl(\var^2-x^2\bigr)}{Z_{L,N}}\exp\Bigl(\frac{L}{2}\epsilon_\alpha x^2\Bigr),
\end{align*}
with $\epsilon_\alpha=E_\alpha/E_{\rm max}$ and
\[
Z_{L,N} = \int_{0}^\infty dx Z_{L,N-1}\bigl(\var^2-x^2)\exp\Bigl(\frac{L}{2}\epsilon_\alpha x^2\Bigr),
\]
where it is assumed that $Z_{L,N}$ corresponds to the system with one single mode at $\epsilon_\alpha$ added to the SD~(\ref{eq:sd}),
while $Z_{L,N-1}$ corresponds to the SD~(\ref{eq:sd}) alone. Let us call
\[
V_{\rm ex} = \var^2-\var_c^2,
\]
the ``excess of variance'' in the system. We get for the condensate the following behavior
\[
p_{\hsigma_\alpha}(x) \propto W_\gamma\Bigl(L^{\frac{\gamma}{1+\gamma}}(x^2-V_{\rm ex}),L^{\frac{1}{1+\gamma}}(1-\epsilon_\alpha)\Bigr)
\]
with now
\[
W_\gamma(x,y) = e^{\frac{-xy}{2}}\int_0^\infty du\ e^{-b_2u^{\gamma+1}}
\cos\Bigl(\frac{xu}{2}-b_1u^{\gamma+1}\Bigr)
\]
So $W_\gamma(x,y) = e^{-xy/2}V_\gamma(-x/2)$ whose plot is shown on  Figure~\ref{fig:condensate} and which asymptotic behavior for large $x$ is given in Appendix~\ref{app:scaling}.
This help us to determine how many  modes condense and the shape of the distribution along these modes, in the vincinity of the upper boundary of the spectrum  corresponding to $\epsilon_\alpha=1$.
Strictly speaking the bump observed on Figure~\ref{fig:condensate} represents the condensation of a mode only for $y=0$, because as soon as $y$ is strictly positive
$W_\gamma(x,y)\sim_{x\to-\infty}e^{-xy/2}/\vert x\vert^{\gamma+2}$, which means that the contribution of the bump to the distribution is suppressed exponentially by a factor
$\exp\bigl(-L^{\frac{1}{\gamma+1}}V_{\rm ex}y/2\bigr)$ by comparison to contributions near $\sigma_\alpha^2=0$. Still  we see that the bump is present for some values $y\in[0,y_\gamma[$.
To know to which modes this corresponds to, 
first note that $y = L^{\frac{1}{1+\gamma}}(1-\epsilon)$ is actually a measure of the rank from the top of the spectrum. Given the SD~(\ref{eq:sd}) the $k$th mode is actually located
in the range $y\in[y_\gamma^{(k)},y_\gamma^{(k+1)}[$
with
\[
y_k \egaldef \Bigl[k\frac{\gamma(\gamma+1)\pi}{\sin(\gamma\pi)}\Bigr]^{\frac{1}{\gamma+1}},
\]
corresponding to a value of $\epsilon$ s.t. $L\int_\epsilon^1du\rho(u) = k$ for large $L$ and finite $k$.
It can be checked numerically that $y_\gamma$ is always below $y_\gamma^{(1)}$ for $\gamma\in]0,1[$, which
means that the bump concerns only the highest mode.

For modes which are detached above the bulk we have to consider the situation with $y=L^{\frac{1}{1+\gamma}}(1-\epsilon)<0$. In that case $p_{\hsigma_\alpha}(x)$ 
has a Bell shape centered around $x^2=V_{\rm ex}$, and according to the asymptotic behavior of $V_\gamma$ given in appendix, 
$p_{\hsigma_\alpha}(x)$ decays like $1/\vert V_{\rm ex}-x^2\vert^{\gamma+2}$ when $x\to 0$ and like $\exp\Bigl(-c_2L(x^2-V_{\rm ex})^{\frac{\gamma+1}{\gamma}}+\frac{L}{2}(x^2-V_{\rm ex})(\epsilon-1)\Bigr)$
for $x^2\gg V_{\rm ex}$ ($(\gamma+1)/\gamma$ is always greater that $1$ for $\gamma\in]0,1[$).

\section{Doubly degenerate spherical RBM} \label{sec:exactA}
Instead of using the saddle point approximation we remark that closed-form expressions of the partition function and of the LL
can be obtained in the case where the spectrum of the weight matrix has discrete levels with a multiplicity of $2$ per level. Defining an RBM obeying this constraint
can be done simply by a duplication of both the input and hidden layer with two identical blocks of the weight matrix but keeping one single spherical constraint on the hidden layer. 
\begin{figure}[ht]
\centerline{\resizebox*{0.7\textwidth}{!}{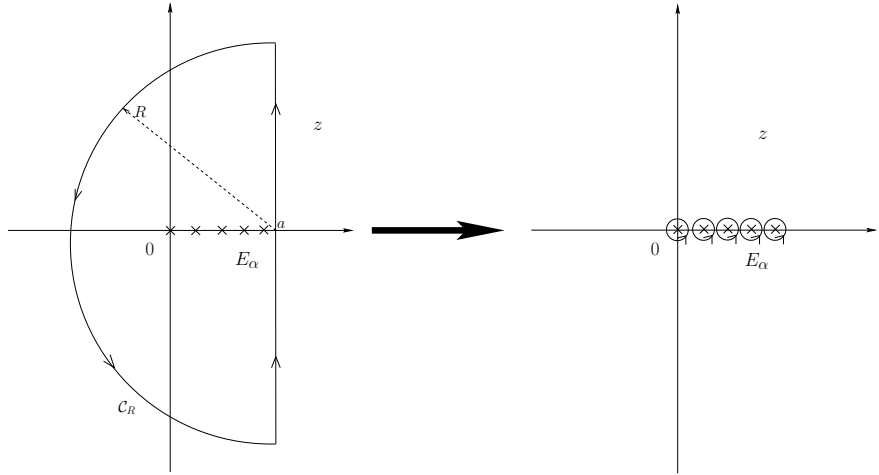}}
\caption{\label{fig:contour} Integration contour deformation.}
\end{figure}
\subsection{Dual formulation}
Let  $K=\max(0,(N_h-N_v)/2)$ and let now $N = \min (N_v,N_h)/2$ represent half the rank of W.
The weight matrix takes now the form:
\[
W = \sum_{\alpha=1\atop\omega\in\{1,2\}}^N w_\alpha u^{\alpha,\omega}v^{\alpha,\omega}.
\]
Given this we then have the following form for the partition function:
\[
Z = \frac{1}{2i\pi}\int_{a-i\infty}^{a+i\infty} \frac{dz}{z^K} \prod_{\alpha=1}^N(z-E_\alpha)^{-1} e^{L\phi(z)}
\]
with now
\[
\phi(z) = \frac{\var^2 z}{2}+\frac{h_0^2}{2z}+\frac{1}{2}\sum_{\alpha=1}^N \frac{h_\alpha^2}{z-E_\alpha},
\]
with 
\[
h_\alpha^2 = h_{\alpha,1}^2+h_{\alpha,2}^2
\]
and 
\begin{align*}
h_{\alpha,\omega} &= w_\alpha\eta_{\alpha,\omega}+\theta_{\alpha,\omega},\qquad \alpha=1,\ldots N\\[0.2cm]
h_0^2 &= \sum_{\alpha=N+1}^{N+K} \bigl(\theta_{\alpha,1}^2+\theta_{\alpha,2}^2\bigr).
\end{align*}
To evaluate $Z$ we can deform the integration contour to the half circle ${\cal C}_R$ as shown on Figure~\ref{fig:contour}:
\[
Z = \lim_{R\to\infty} \frac{1}{2i\pi}\oint_{{\cal C}_R} \frac{dz}{z^K} \prod_\alpha(z-E_\alpha)^{-1} e^{L\phi(z)},
\]
thanks to the following bound for the contribution on the half circle for $R$ sufficiently large:
\[
 R\left\vert z^{-K}\prod_\alpha(z-E_\alpha)^{-1}e^{L\phi(z)} \right\vert \le  \frac{A}{R^{N_h-1}} \to_{R\to\infty} 0.
\]
Then, since the integrand is holomorphic  everywhere inside the domain enclosed by ${\cal C}_R$ except on the singularities $z = E_\alpha$, 
we can deform the contour as shown on Figure~\ref{fig:contour}
in terms of small anti-clockwise circles ${\cal C}_{\epsilon,\alpha}$ of radius $\epsilon$ around each singularities including $z=0$ for ${\cal C}_{\epsilon,0}$, $\epsilon$
being small enough such that each ${\cal C}_{\epsilon,\alpha}$ encloses one single singularity. $Z$ is then expressed as
\[
Z = \frac{1}{2i\pi}\sum_{\alpha=0}^N\oint_{{\cal C}_{\alpha,\epsilon}} \frac{dz}{z^K}\prod_\beta(z-E_\beta)^{-1} e^{L\phi(z)},
\]
which after expanding for each contour the enclosed singular part in the exponential reads
\[
Z = \sum_{n=0}^\infty\Bigl( \frac{\bigl(L h_0^2\bigr)^{n}}{4^n n!^2}f_0^{(n+K-1)}(0)+
\sum_{\alpha=1}^N  \frac{\bigl(L h_\alpha^2\bigr)^{n}}{4^n n!^2}f_\alpha^{(n)}(0)e^{\frac{L}{2}\var^2 E_\alpha}\Bigr),
\]
where
\begin{align*}
f_0(z) &= \prod_{\beta}\frac{1}{z-E_\beta}\exp\Bigl(\frac{L}{2}\Bigr[\var^2 z+
\sum_{\beta}\frac{h_\beta^2}{z-E_\beta}\Bigr]\Bigr),\\[0.2cm]
f_\alpha(z) &= \frac{1}{(z-E_\alpha)^K}
\prod_{\beta\ne\alpha}\frac{1}{z+E_\alpha-E_\beta}\exp\Bigl(\frac{L}{2}\Bigr[\var^2 z+\frac{h_0^2}{z+E_\alpha}+
\sum_{\beta\ne\alpha}\frac{h_\beta^2}{z+E_\alpha-E_\beta}\Bigr]\Bigr).
\end{align*}
It is rather tedious to write down this expression for $Z$ more explicitly, instead at this point let us consider the case when
 all $h_\alpha=0$.
Then we simply get
\begin{equation}\label{eq:Z}
Z = \sum_{\alpha=1}^N \frac{1}{E_\alpha^K}
\Bigl(\exp\bigl(\frac{\var^2 LE_\alpha}{2}\bigr)-\sum_{k=0}^{K-1}\frac{1}{k!}\bigl(\frac{\var^2 LE_\alpha}{2}\bigr)^k\Bigl) 
\prod_{\beta\ne\alpha}(E_\alpha-E_\beta)^{-1}.
\end{equation}
Reducing this expression to the same denominator leads us to express the partition function as a ratio of two determinants:
\[
Z = \frac{
\left(
\begin{matrix}
1 & E_1 & E_1^2 &\ldots & E_1^{N-2} & \sum_{k=N-1}^\infty\frac{1}{(k+K)!}\bigl(\frac{\var^2 LE_1}{2}\bigr)^k \\
1 & E_2 & E_2^2 &\ldots & E_2^{N-2} & \sum_{k=N-1}^\infty\frac{1}{(k+K)!}\bigl(\frac{\var^2 LE_2}{2}\bigr)^k \\
\ldots & \ldots &\ldots &\ldots &\ldots & \ldots \\[0.2cm]
1 & E_{N} & E_{N}^2 &\ldots & E_{N}^{N-2} & \sum_{k=N-1}^\infty\frac{1}{(k+K)!}\bigl(\frac{\var^2 LE_{N}}{2}\bigr)^k 
\end{matrix}
\right)
}
{
\left(
\begin{matrix}
1 & E_1 & E_1^2 &\ldots & E_1^{N-2} & E_1^{N-1} \\
1 & E_2 & E_2^2 &\ldots & E_2^{N-2} & E_2^{N-1} \\
\ldots & \ldots &\ldots &\ldots &\ldots & \ldots \\[0.2cm]
1 & E_{N} & E_{N}^2 &\ldots & E_{N}^{N-2} & E_{N}^{N-1} 
\end{matrix}
\right)
}
\]
This ratio is actually a weighted sum of particular Schur polynomials (see e.g.~\cite{Sagan}), each one being a positive symmetric function of the 
energy levels $E_\alpha$. There is a generating function for these thanks the following identity
\[
\left(
\begin{matrix}
1 & E_1 & E_1^2 &\ldots & E_1^{\tiny N-2} & \frac{1}{1-tE_1} \\
1 & E_2 & E_2^2 &\ldots & E_2^{\tiny N-2} & \frac{1}{1-tE_2}\bigr) \\
\ldots & \ldots &\ldots &\ldots &\ldots & \ldots \\[0.2cm]
1 & E_{\tiny N} & E_{\tiny N}^2 &\ldots & E_{\tiny N}^{\tiny N-2} & \frac{1}{1-tE_{\tiny N}}\bigr) 
\end{matrix}
\right)
= \prod_{\alpha <\beta}(E_\alpha-E_\beta)\frac{t^{\tiny N-1}}{\prod_\alpha(1-tE_\alpha)},
\]
we have
\begin{equation}\label{eq:Zqueues}
Z = \sum_{\{n_\alpha\}}\frac{\bigl(\frac{\var^2 L}{2}\bigr)^{K+N-1}}{(\sum_\alpha n_\alpha+K+N-1)!}\prod_{\alpha=1}^{N}\bigl(\frac{L}{2}\var^2 E_\alpha\bigr)^{n_\alpha},
\end{equation}

\noindent where the $n_\alpha$'s run over $\mathbbm N$.

\subsection{Urn model interpretation}
Expression~(\ref{eq:Zqueues})  allows us to make the connection with another type of models studied in statistical physics, namely 
urn models (see e.g.~\cite{GoLu}) which generalize the Ehrenfest model to an extensive number of urns. These are 
in fact a special case of queuing network processes~\cite{GePu} well studied in probability theory.
In the queuing theory, general class of queuing networks have been identified which have simple explicit steady state measures~\cite{BCMP}. The first one is the so called
Jackson network with exponential service rates~\cite{Jackson} possibly open or closed.
Up to a multiplicative constant, the form~(\ref{eq:Zqueues}) coincides with the normalization of the invariant measure of a closed Jackson network of queues.
Here each index $\alpha$ refers to a queue characterized by a service rate $\mu_\alpha$ and $n_\alpha$ is interpreted as the number of clients in the queue.
These queues are assembled into a network by fixing a set of routing probabilities among them. Many different configurations can possibly lead to the same measure corresponding to~(\ref{eq:Zqueues}). 
Let us give an instance of such a network fulfilling this constraint. In addition of the queues already defined we add an additional one with index $\alpha=0$ corresponding  to the reservoir,
$n_0$ being its number of clients supposedly large.  
The routing is then simply defined by the given ordering of the queues: a client in queue $\alpha$ is routed to queue $(\alpha+1)\ mod(N)$, which means that the system is closed and the
queues are arranged along a circle. Let $N_\infty = n_0+\sum_{\alpha>0}n_\alpha$ represent the total number of clients, considered arbitrarily large.
Consider the following service rates:
\[
\mu_\alpha =
\begin{cases}
\DD  \hspace{1.5cm} E_\alpha^{-1} \hspace{2.2cm} \forall\alpha\in\{1,\ldots N\},\\[0.2cm]
\DD  \frac{\var^2 L}{2(N_\infty-n_0+K+N-1)}\qquad\text{for}\ \alpha=0.
\end{cases}
\]
We are then in the conditions of the Jackson theorem, namely that each service rate depends only the state of the corresponding queue and that there exists a set of actually constant
arrival rates $\lambda_\alpha= \lambda$, satisfying the so-called traffic equation (i.e. flux conservation on the network). Hence the corresponding measure has  the following form
\[
P({\bm n}) = \frac{1}{\tilde Z(N_\infty)}\prod_{n=0}^{n_0-1}\frac{2\lambda(N_\infty-n+K+N-1)}{\var^2 L}
\prod_{\alpha=1}^N \Big(\frac{\lambda}{\mu_\alpha}\Big)^{n_\alpha}\delta\Bigl(N_\infty-\sum_{\alpha=0}^N n_\alpha\Bigr),
\]
where $\lambda$ is arbitrary, with (after re-arranging the summation)
\[
\tilde Z(N_\infty) = \lambda^{N_\infty}\sum_{\{n_\alpha,\alpha>0\}}\ind{\sum_{\alpha}n_\alpha\le N_\infty}
\Bigl(\frac{L\var^2}{2}\Bigr)^{\sum_\alpha n_\alpha-N_\infty}
\frac{(N_\infty+K+N-1)!}{\bigl(\sum_\alpha n_\alpha+K+N-1\bigr)!}\prod_{\alpha=1}^N E_\alpha^{n_\alpha}.
\]
So if we now let the size of the reservoir (controlled by $N_\infty$) become  sufficiently large, 
we see that up to an irrelevant $N_\infty$-dependent factor, $\tilde Z(N_\infty)$ coincides with $Z$:
\[
Z = 
\lim_{N_\infty\to\infty} \frac{\lambda^{-N_\infty}(\var^2 L/2)^{N_\infty+N+K-1}}{(N_\infty+N+K-1)!}\ \tilde Z(N_\infty).
\]

\noindent From the practical point of view, $Z$ can be computed with arbitrary precision, thanks to the following recursion.
Let
\[
Z_{l,m}[E_1,\ldots,E_l] \egaldef \sum_{n_1,\ldots,n_l}\ind{\sum_{\alpha=1}^l n_\alpha=m} \prod_{\alpha=1}^l E_\alpha^{n_\alpha}.
\]
We have
\begin{equation}\label{eq:Zrec}
Z = \sum_{m=0}^\infty \frac{\bigl(\frac{\var^2 L}{2}\bigr)^{m+K+N-1}}{(m+K+N-1)!} Z_{N,m}\bigl[E_1,\ldots,E_{N}\bigr].
\end{equation}
In order to compute $Z$ numerically we make use of the following recursion:
\[
Z_{l+1,m}[E_1,\ldots,E_{l+1}] = \sum_{k=0}^mE_{l+1}^k Z_{l,m-k}[E_1,\ldots,E_l].
\]
Thanks to this recursion, if now we fix an upper bound $M = O(L)$ (in the condensed phase) of the maximal number of clients in order to reach a given precision for $Z$, we end up with a complexity
${\cal O}(L^2)$ to estimate the partition function.
\begin{figure}
\centering
\includegraphics[scale=0.7]{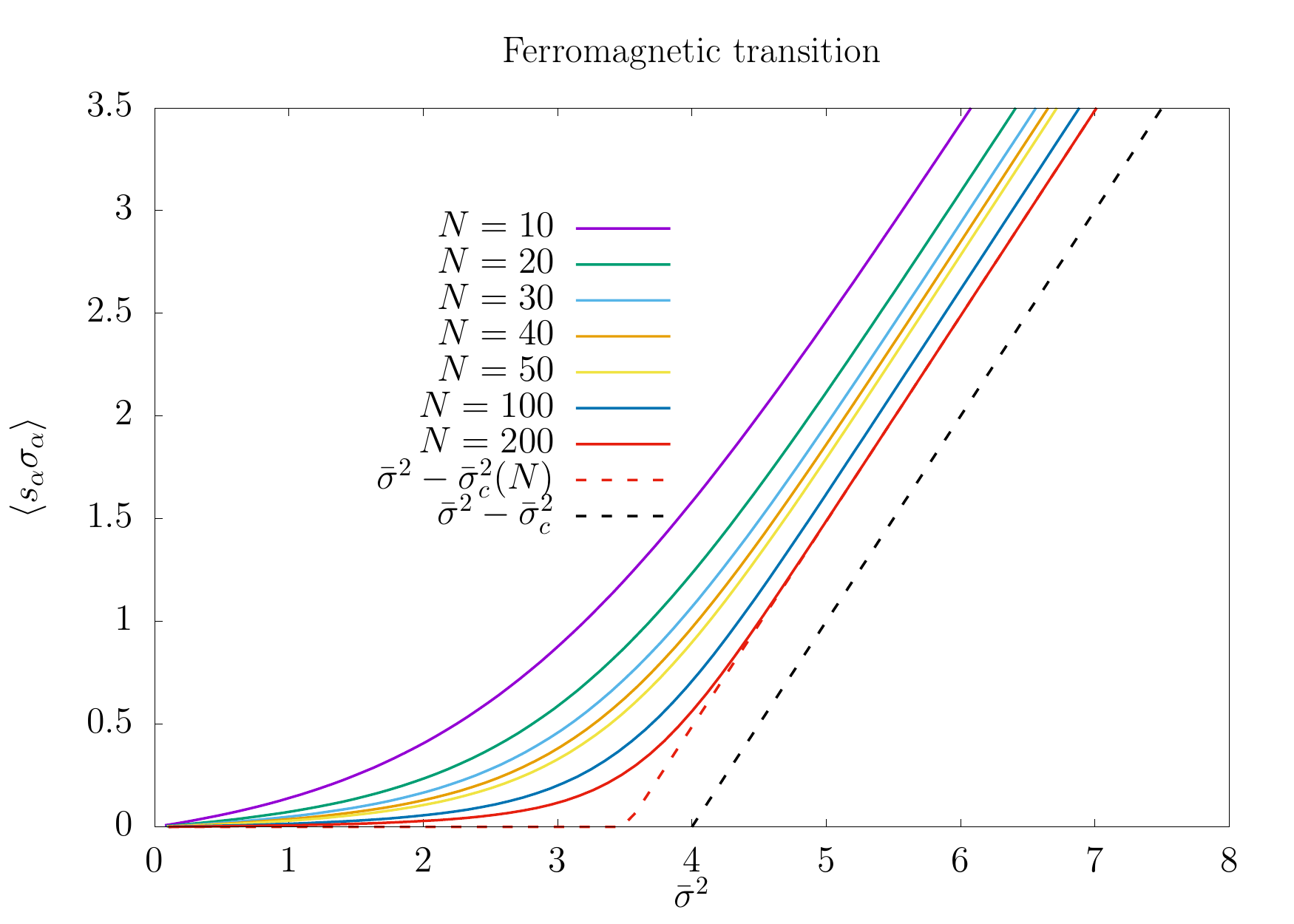}
\caption{Two-points response function $\langle\hs_\alpha\hsigma_\alpha\rangle$ with $\alpha$ corresponding to the highest mode, obtained with help of~(\ref{eq:Zrec}),
  of an RBM having the SD~(\ref{eq:sd}) with $\gamma=0.5$ and $\kappa=1$. The finite size dependence of the critical $\var^2_c$ is clearly pronounced. 
  The red dashed curve indicates the finite size behavior when $N=200$.
  \label{fig:response}}
\end{figure}
On Figure~\ref{fig:response} is shown the finite size dependence of the two-point function $\langle \hs_\alpha\hsigma_\alpha\rangle$ using these recursions. 
\subsection{The ferromagnetic transition as a Bose-Einstein condensation}
It is known for a very long time that the spherical model is related to the ideal Bose gas and that the transition is analogous 
to the Bose-Einstein condensation (see~\cite{Pastur} and references herein). In this queueing process language we can make it very explicit.
In its original formulation, the ferromagnetic transition is associated to sharp increase of the magnetization projected on the dominant mode,
which results for the thermal expectation $\langle\hs_\alpha^2\rangle_{\rm RBM}$ along this mode, in a change of from a ${\cal O}(1/L)$ to a ${\cal O}(1)$ behavior.
In the queuing terminology, leaving aside the queue corresponding to the reservoir, we expect to see a transition where the queue associated to the smallest service rate
absorbs a finite fraction of the total number of clients ${\cal O}(L)$ present in the system (not in the reservoir). In fact the two are closely related since we have:
\begin{align*}
\langle\hs_\alpha^2\rangle_{\rm RBM} &= \frac{2v_\alpha^2}{L}\frac{\partial \log(Z)}{\partial v_\alpha}\\[0.2cm]
\langle n_\alpha\rangle_{\rm RBM} &= E_\alpha\frac{\partial \log(Z)}{\partial E_\alpha}
\end{align*}
if $v_\alpha$ is the prior variance of mode $\hs_\alpha$ set to the default value $v_\alpha=1$ in~(\ref{eq:Erbm}).
We then have the relationship (in absence of external fields)
\begin{align*}
\langle n_\alpha\rangle_{\rm RBM} &= L\langle\hs_\alpha\hsigma_\alpha\rangle_{\rm RBM}\\[0.2cm]
&= \frac{L}{2}\bigl(\langle\hs_\alpha^2\rangle_{\rm RBM}-\frac{1}{L}\bigr)\ge 0. 
\end{align*}

\noindent The phase transition identified previously actually correspond to an ordinary Bose-Einstein condensation when reinterpreting the $n_\alpha$ 
as occupation numbers of states $\alpha$ of energy $\epsilon_\alpha$ in this last expression, with the identification
\[
e^{-\beta\epsilon_\alpha} = E_\alpha
\]
and the fugacity $\nu$ representing
\[
e^{-\beta\nu} = \frac{2z}{\var^2 L}.
\]
Then the critical value $\var_c$ of $\var$ previously given in (\ref{eq:varc}) is reintepreted as
\[
L\var^2_c = \int_{\epsilon_{min}}^{+\infty} d\epsilon \frac{\rho(\epsilon)}{\exp\bigl[\beta(\epsilon-\nu)\bigr]-1}\Bigr\vert_{\nu=\epsilon_{min}},
\]
$L \var^2_c$ corresponding then to the maximum number of bosons that can be inserted into the system without condensing into the ground state
$\epsilon_{min}$.\\[0.5cm]

\section{Dynamics of learning}\label{sec:dyn}

At present we have all the material to setup an ``exact'' learning method of the Gaussian-spherical RBM, i.e. based on exact response function. 
The continuous learning equations (\ref{eq:wa},\ref{eq:hsa},\ref{eq:omega}) given  in Section~\ref{sec:LA} are integrated straightforwardly.
\begin{figure}
	\centering
	\includegraphics[scale=0.9]{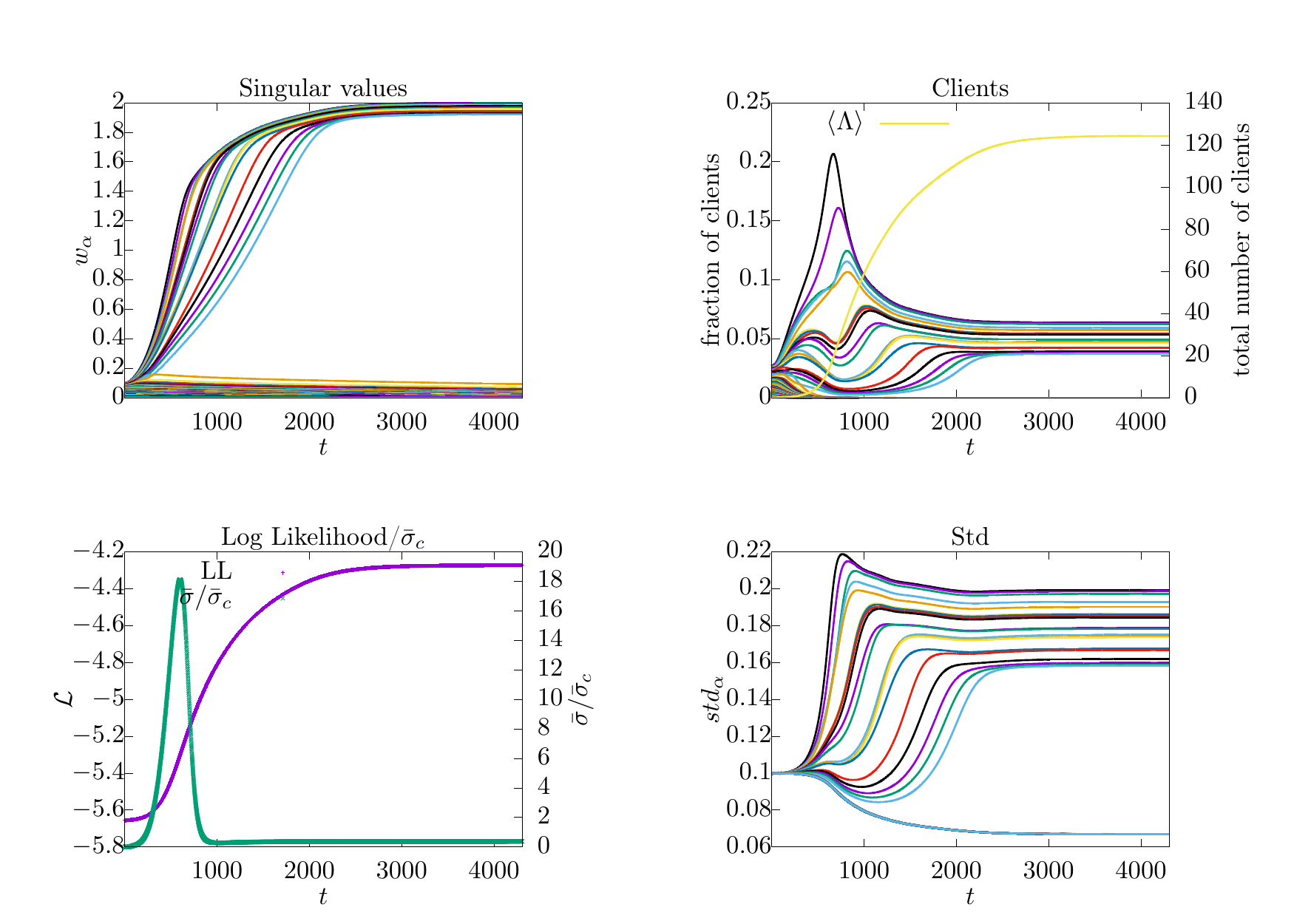}
\caption{Learning dynamics of an RBM of size $(N_v,N_h)=(100,100)$ learned on a synthetic dataset of $2000$ samples distributed in the neighborhood of a $20$-d 
ellipsoid embedded into a $100$-d space. The dynamics of the singular values is shown on the top left panel. Correspondingly the evolution of 
the number of clients filling the queues is shown on the top right panel or equivalently the variance along each mode on the bottom right. 
Here all the modes corresponding to the principal axes of the ellipsoid do eventually condense.\label{fig:learning}}
\end{figure}
All the one and two-point correlation functions involved in these equations can be estimated with arbitrary precision in principle, from the previous section. As a result
we can generate the deterministic learning trajectories shown on Figure~\ref{fig:learning}. The synthetic data used to train this RBM are generated
from a distribution which support is localized in the neighborhood of an ellipsoid of small dimension embedded into a larger dimensional space. As can be checked 
the modes which emerge eventually align with one of the principal axes of the ellipsoid. The order of arrival is in correspondence with 
the order of the values of the corresponding principal axes the modes are aligning with. Here the needed time to condense combines the time it takes to align in the right direction
and the amplification time of the singular value itself. 
\begin{figure}
\centering
\includegraphics[scale=0.7]{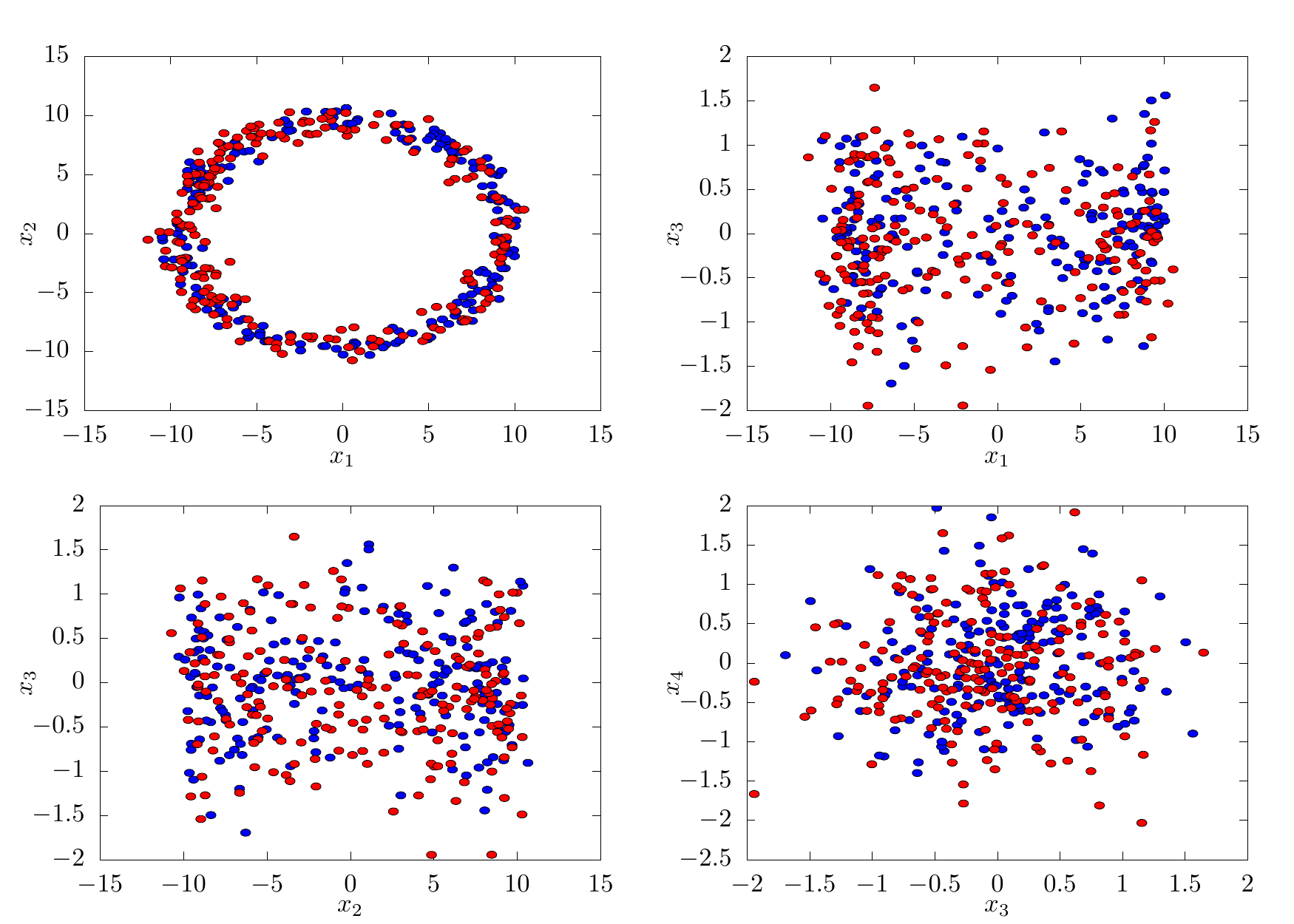}
\caption{Scatter plot of the training data (blue) and sampled data from the learned RBM (red) projected on the first four svd modes of the weight matrix, for a problem
  in dimension $N_v=50$ with two condensed modes.
\label{fig:scatter1}}
\end{figure}

\begin{figure}
\centering
\includegraphics[scale=0.7]{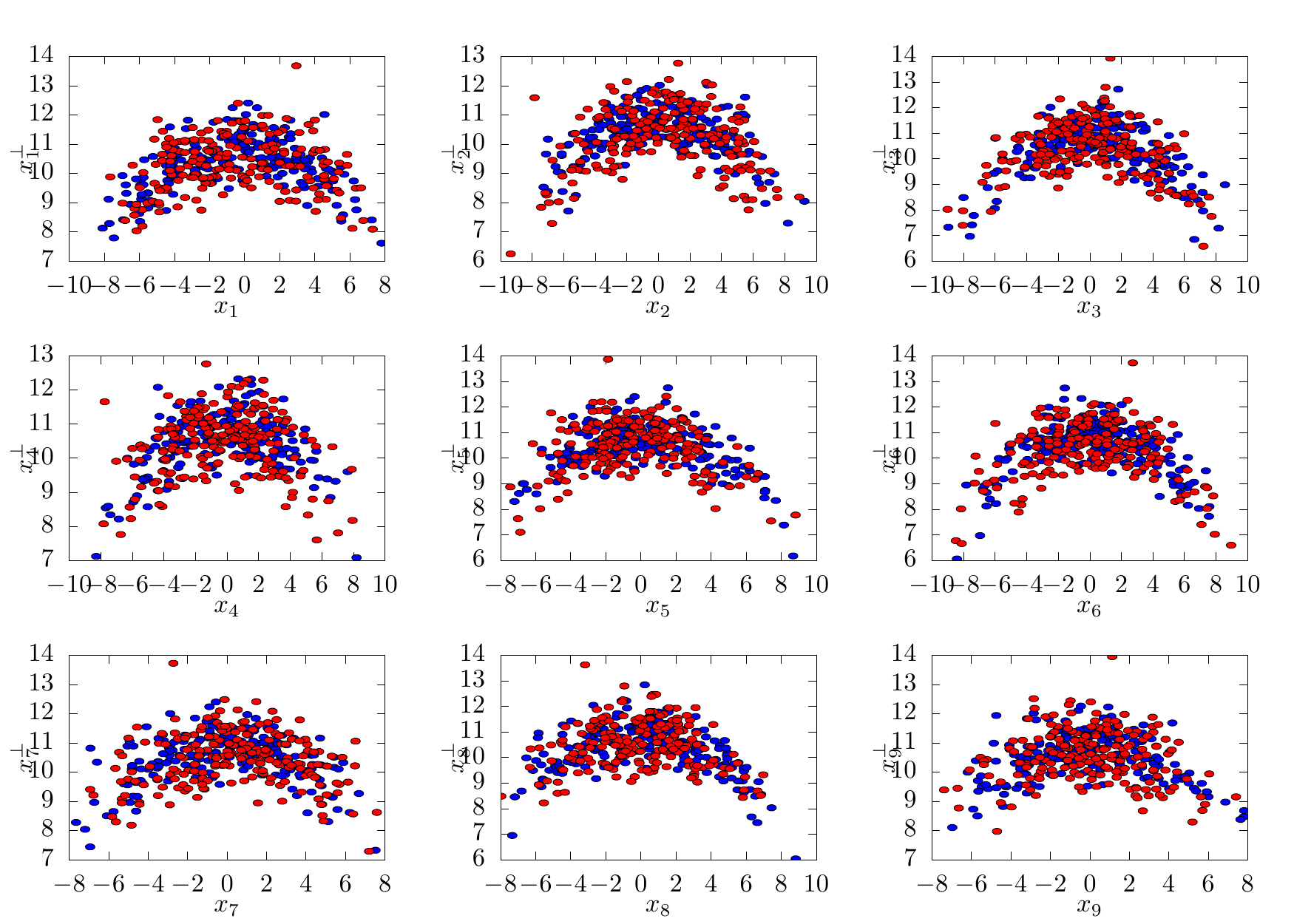}
\caption{Scatter plot of the training data (blue) and sampled data from the learned RBM (red) projected on the first nine svd modes of the weight matrix ($x_i$) against the norm of the
  the orthogonal complement ($x_i^\perp = \sqrt{\vert x\vert^2-x_i^2}$), for a problem in dimension $N_v=60$ with nine condensed modes.
\label{fig:scatter2}}
\end{figure}
The scatter plots shown on Figures~(\ref{fig:scatter1},\ref{fig:scatter2}) illustrates the ellipsoid shape of the distribution when more than one mode get condensed. 

With respect to the thermodynamic analysis given in Section~\ref{sec:condensate}, the final state of the RBM found by the learning process do not correspond to 
a continuous bulk of singular values with some given exponent $\gamma$. Instead it is a situation where a small number of modes is detached from the bulk of the SD, 
with mutual distances of order $1/L$, all condensing and capturing in almost equal proportion a finite fraction of the overall variance. 

\section{Perspectives}
The RBM model presented in Section~\ref{sec:exactA} is clearly useful only from the theoretical point of view, in particular it could serve to test some heuristic learning strategies.
Interestingly it is modelling  distribution with a continuous manifold (ellipsoid or portion of ellipsoid if biases are switched on) 
when more than one mode get condensed. By contrast an RBM with binary latent 
variables tends to form small spherical clusters to cover the training dataset. This property could maybe used in a more complex setting.   
Indeed, a more general model which can be still solvable using asymptotic response functions of Section~\ref{sec:asymp} is the following.
First we remark that if we partition the hidden variables into $P = N_h/n$ subsets of size $n$, and define a spherical constraint on each of these
subset, we thereby define a family of models which interpolates between the Gaussian RBM considered so far and an RBM with Ising latent variables when varying $n$
between $N_v$ and $1$. Then consider an RBM with $N_v$ and $P$ fixed, but with varying sizes of the partition $\{n_q^h,q=1,\ldots P\}$
with $\sum_q n_q=P$. For the visible variables we consider first an arbitrary basis $\uu$ partitioned into $P$ subspaces again of sizes $\{n_q^v,q=1,\ldots P\}$,
the overall dimension of the visible space being $N_v=\sum_q n_q^v$. We can then define an RBM expressed as a direct product of smaller RBM on this partition directly
in the svd mode representation of the global weight matrix:
\[
P(\hs,\hsigma) = \prod_{q=1}^P\frac{1}{Z_q}e^{-L E_q( \bm{\hs},\bm{\hsigma})}\delta\Bigl( \sum_{\alpha=1}^{n_q^h} \sigma_{q,\alpha}^2 - n_q^h \Bigr)
\]
where now
\[
E_q( \bm{\hs},\bm{\hsigma}) = \sum_{\alpha=1}^{\min(n_q^v,n_q^h)}\bigl[w_{q,\alpha}\hs_{q,\alpha}\hsigma_{q,\alpha}+\eta_{q,\alpha}\hs_{q,\alpha}+\theta_{q,\alpha}\hsigma_{q,\alpha}\bigr]
-\frac{1}{2}\sum_{\alpha=1}^{n_q^v}\frac{\hs_{q,\alpha}^2}{\var_{q,\alpha}^2},
\]
($\var_{q,\alpha}$ being default variances), and each partition function $Z_q$ can be computed by saddle points approximations.
The matrix $\uu$ as well as the fields $\heta$ and $\htheta$  has
to be learned while $\uv$ is predefined with each mode $(q,\alpha)$ localized on the corresponding subset of hidden variables corresponding to the
$q$th partition.

\vspace{1cm}
\bibliographystyle{unsrt}
\bibliography{rbm}

\begin{appendices}

\section{Response functions}\label{app:resp}
All response functions at leading order are expressed as derivatives  w.r.t. external fields of    
$\phi$ given in~(\ref{eq:phi}) taken at the saddle point $z_0$. 
We have
\begin{align*}
\langle\hs_\alpha\rangle &= \frac{\partial\phi}{\partial\heta_\alpha}(z_0)\\[0.2cm]
\langle \hsigma_\beta\rangle &= \frac{\partial\phi}{\partial\theta_\beta}(z_0)\\[0.2cm]
\langle\hs_\alpha\hsigma_\beta\rangle &= \frac{\partial\phi}{\partial\heta_\alpha}(z_0)\frac{\partial\phi}{\partial\theta_\beta}(z_0)
+\frac{1}{L}\frac{\partial^2\phi}{\partial\heta_\alpha\partial\theta_\beta}(z_0).
\end{align*}
We get these as a function of $z_0$:
\begin{align*}
\langle\hs_\alpha\rangle &= \Bigl(\heta_\alpha+\frac{w_\alpha h_\alpha}{z_0-w_\alpha^2}\Bigr),\\[0.2cm]
\langle \hsigma_\beta\rangle &= \frac{h_\alpha}{z_0-w_\alpha^2},\\[0.2cm]
\langle\hs_\alpha\hsigma_\beta\rangle - \langle\hs_\alpha\rangle \langle \hsigma_\beta\rangle &= 
\frac{1}{L}w_\alpha\frac{\partial }{\partial\theta_\beta}\Bigl(\frac{h_\alpha}{z_0-w_\alpha^2}\Bigr)\\[0.2cm]
&= \frac{w_\alpha}{L}\Bigl(\frac{\delta_{\alpha\beta}}{z_0-w_\alpha^2}-\frac{h_\alpha}{(z_0-w_\alpha^2)^2}\frac{\partial z_0}{\partial \theta_\beta}\Bigr)\\[0.2cm]  
&= \frac{w_\alpha}{L}\left(\frac{\delta_{\alpha\beta}}{z_0-w_\alpha^2}-\Bigl(\sum_\gamma\frac{h_\gamma^2}{(z_0-w_\gamma^2)^3}-\frac{1}{L}\frac{1}{(z_0-w_\gamma^2)^2}\Bigr)^{-1}
\frac{h_\alpha h_\beta}{(z_0-w_\alpha^2)^2(z_0-w_\beta^2)^2}\right),
\end{align*}
where $\partial z_0/\partial\htheta_\alpha$ is obtained from the saddle point condition.

\section{Asymptotic expressions for the condensate}\label{app:scaling}
The large deviation function~(\ref{eq:phi}) reads (after dropping irrelevant terms) in the continuous formulation
\[
\phi(z) = \frac{1}{2}\var^2z+\frac{1}{2}\int_0^{E_{\rm max}}dE \rho(E)\Bigl(\frac{h(E)^2}{z-E}-\log(z-E)\Bigr).
\]
First we make the change of variable $(z-E_{\rm max})/E_{\rm max} \to z$ in the integral representation~(\ref{eq:Zcontour})
and change accordingly the definition of the spectral density $\rho(E)dE \to \rho(u=E/E_{\rm max})du$ and similarly for $\rho(E)h(E)^2$,
while $E_{\rm max}$ is absorbed in the definition of $\var^2E_{\rm max} \to \var^2$. This leads then to 
\[
Z(\var,h) = \frac{E_{\rm max}}{2i\pi}\int_{-i\infty}^{i\infty}dz\ e^{\frac{L}{2}\phi(z,\var,h)},
\]
with 
\[
\phi(z,\var,h) = \var^2(z+1)-\int_0^1 du \rho(u)\Bigl(\frac{h(u)^2}{z+1-u}-\log(z+1-u)\Bigr).
\]
From the expressions~(\ref{eq:sd},\ref{eq:h}) and the definition of hypergeometric functions we have
\[
\partial_z \phi(z,\var,h) = \var^2 -  h^2\Bigl(\frac{1}{z+1}\Bigr)^2F\Bigl(2,1+\beta;2;\frac{1}{z+1}\Bigr) -
\frac{\kappa}{2}\frac{1}{z+1}F\Bigl(1,1-\gamma;2;\frac{1}{z+1}\Bigr).
\]
More general beta distributions with arbitrary exponents would lead as well to  hypergeometric functions with different parameters.
Our choice leads to more explicit expressions. Indeed,  
using hypergeometric transformations formulas (See Gradshteyn \& Ryzhik) we have:
\begin{align*}
  F(2,1+\beta;2;u) &= (1-u)^{-1-\beta}F(0,1-\beta;2;u)  \\[0.2cm]
  &= (1-u)^{-1-\beta} \\[0.2cm]
  F(1,1-\gamma;2;u) &= (1-u)^\gamma F(1,1+\gamma;2;u) \\[0.2cm]
  &= \frac{1}{\gamma u}\Bigl[1-(1-u)^\gamma\Bigr]
\end{align*}
So we finally get for the saddle point equation
\begin{align*}
\phi'(z) &= \var^2-h^2 \frac{(z+1)^{\beta-1}}{z^{1+\beta}} +\var_c^2\Bigl[\Bigl(\frac{z}{z+1}\Bigr)^\gamma-1\Bigr],\qquad\text{for}\qquad\gamma\ge 0\\[0.2cm]
&= 0,
\end{align*}
with
\[
\var_c^2 = \frac{\kappa}{\gamma}.
\]
Upon integration over $z$ we obtain the expression (\ref{eq:phizvarh}) of $\phi(z,\var,h)$. The expression of the partition function in term of the scaling 
function $V_{\gamma,\beta}$ is obtained after the rescaling $L^{1/(1+\gamma)}z \to z$.

In absence of external field it becomes obvious that for $\var_c<\var$ there is no saddle point solution in the domain $z\ge 0$. This situation has been 
analyzed in depth in a slightly different context of condensation in zero range processes~\cite{EvMaZi}.
In that case, the partition function has a scaling behavior  
\[
Z_{L,N}(\var^2) \simeq L^{-\frac{1}{1+\gamma}} V_\gamma\Bigl(L^{\frac{\gamma}{1+\gamma}}(\var^2-\var_c^2)\Bigr),
\]
given in terms of the scaling function (slightly adapting the notation of~\cite{EvMaZi})  
\begin{align*}
V_\gamma(x) &= \frac{1}{2i\pi}\int_{-i\infty}^{i\infty} du e^{ux+bu^{1+\gamma}},\\[0.2cm]
&= \frac{1}{\pi}\int_0^\infty e^{-b\sin(\gamma\pi/2)u^{\gamma+1}}\cos\Bigl(b\cos(\gamma\pi/2)u^{\gamma+1}-ux\Bigr).
\end{align*}
Here $L(\var^2-\var_c^2) = V_{\rm ex}$ represents the excess of variance that forces the system to condense on the highest modes. The asymptotic behaviour 
of $V_\gamma$ studied in~\cite{EvMaZi} rewrites here:
\[
V_\gamma(x) =
\begin{cases} 
\DD \frac{b\gamma(\gamma+1)}{\Gamma(1-\gamma) x^{\gamma+2}}  \hspace{1.8cm}\text{as}\qquad x\to\infty \\[0.2cm]
\DD \frac{b^{-\frac{\gamma}{\gamma+1}}}{(\gamma+1)\Gamma\bigl(\frac{\gamma}{\gamma+1}\bigr)}  \hspace{1cm}\text{as}\qquad x=0\\[0.4cm]
\DD c_1 \vert x\vert^{\frac{1-\gamma}{2\gamma}} \exp\bigl(-c_2 \vert x\vert^\frac{\gamma+1}{\gamma}\bigr)  \qquad\text{as}\qquad x\to-\infty
\end{cases}
\]
with 
\[
c_1 = \frac{1}{\sqrt{2\pi\gamma}\bigl(b(\gamma+1)\bigr)^{\frac{1}{2\gamma}}} \qquad\text{and}\qquad
c_2 = \frac{\gamma}{\gamma+1}\bigl(b(\gamma+1)\bigr)^{-\frac{1}{\gamma}},
\]
leading to 
\[
Z_{L,N}(\var^2) \propto \frac{L}{V_{\rm ex}^{\gamma+2}} 
\]
in the regime $V_{\rm ex} = {\cal O}(L)$.

\end{appendices}
\end{document}